\documentclass[preprint,5p,times,twocolumn,compress]{elsarticle}
\usepackage[intlimits]{amsmath}
\usepackage{slashed}
\usepackage{amssymb}
\usepackage[T1]{fontenc}
\usepackage[normalem]{ulem} 
\usepackage{color}
\usepackage{mathtools} 
\usepackage{tensor}
\usepackage{icomma}
\usepackage{bbm} 
\usepackage{bbold}
\usepackage{pbox}
\usepackage{simplewick} 
\usepackage{enumitem}
\makeatletter
\def\ps@pprintTitle{%
 \let\@oddhead\@empty
 \let\@evenhead\@empty
 \def\@oddfoot{}%
 \let\@evenfoot\@oddfoot}
\makeatother

\newcommand{\ie}{\emph{i.e.} }		
\newcommand{\eg}{\emph{e.g.} }

\DeclareMathOperator{\tr}{Tr} 
\newcommand{\SU}[1]{\ensuremath{\text{SU(#1)}}}
\newcommand{\rd}{\ensuremath{\mathrm{d}}}

\newcommand{\e}{\ensuremath{\mathrm{e}}}
\newcommand{\I}{\ensuremath{\mathrm{i}}}

\newcommand{\eye}{\ensuremath{\mathbbm{1}}}	
\newcommand{\bra}[1]{\langle #1|}
\newcommand{\ket}[1]{|#1\rangle}
\newcommand{\braket}[2]{\langle #1|#2\rangle}
\newcommand{\expval}[1]{\left\langle #1 \right\rangle}	
\newcommand{\op}{\ensuremath{\mathcal{O}}}
\newcommand{\Xb}{\ensuremath{\overline{X}}}
\newcommand{\Wb}{\ensuremath{\overline{W}}}
\newcommand{\Zb}{\ensuremath{\overline{Z}}}
\newcommand{\It}{\ensuremath{\tilde{I}}}
\newcommand{\Jt}{\ensuremath{\tilde{J}}}

\newcommand{\ua}{\ensuremath{\uparrow}}
\newcommand{\da}{\ensuremath{\downarrow}}
\newcommand{\Y}{\ensuremath{\hat{Y}}}
\newcommand{\Ycl}{\ensuremath{Y^{\cl}}}
\newcommand{\gym}{\ensuremath{g_{\text{YM}}}}
\newcommand{\TT}{\ensuremath{\mathbf{T}}}
\newcommand{\bMPS}{\ensuremath{\bra{\text{MPS}}}}
\newcommand{\kMPS}{\ensuremath{\ket{\text{MPS}}}}
\newcommand{\MPS}{\ensuremath{\text{MPS}}}
\newcommand{\QQ}{\ensuremath{Q}}
\newcommand{\m}{\ensuremath{\mathrm{m}}}
\newcommand{\onecontr}{\text{1 contr.}}
\newcommand{\tree}{\text{tree}}
\newcommand{\Modd}{\tilde{M}}
\newcommand{\cl}{\ensuremath{\text{cl}}}
\newcommand{\fatZero}{\ensuremath{\mathbb{0}}}

\usepackage[hidelinks]{hyperref}

\begin{document}

\begin{frontmatter}



  \title{Two-point functions of SU(2)-subsector and length-two operators in dCFT}
\author{Erik Wid\'en}
\address{Nordita\\
  KTH Royal Institute of Technology and Stockholm University\\
  Roslagstullsbacken 23, SE-106 91 Stockholm, Sweden\\[1mm]
Department of Physics and Astronomy, Uppsala University\\
SE-751 08 Uppsala, Sweden\\[1mm]
\texttt{\upshape erik.widen@nordita.org}
}



\begin{abstract}
  We consider a particular set of two-point functions in the setting of $\mathcal{N}=4$ SYM with a defect, dual to the fuzzy-funnel solution for the probe D5-D3-brane system. The two-point functions in focus involve a single trace operator in the SU(2)-subsector of arbitrary length and a length-two operator built out of any scalars. By interpreting the contractions as a spin-chain operator, simple expressions were found for the leading contribution to the two-point functions, mapping them to earlier known formulas for the one-point functions in this setting.
\end{abstract}

\end{frontmatter}


\section{Introduction}
\makebox[0pt]{\hfill\raisebox{72 ex}[0pt][0pt]{\hspace{0.96\textwidth}\parbox{\textwidth}{ \raggedleft \footnotesize \texttt{NORDITA 2017-034}\\\texttt{UUITP-14/17}}}}%
Integrable structures in $\mathcal{N}=4$ SYM have been explored extensively since they were first noted in \cite{ci:Minahan:2002ve} and have provided a useful tool for both deeper field theoretic understanding and numerous tests of the AdS/CFT correspondence. For a pedagogical overview of the first decade, see \cite{ci:Beisert:2010jr}. Among other directions, the work has lead on to look for, and to employ, surviving integrability in similar theories, departing in different ways from $\mathcal{N}=4$ SYM. One particular branch of this focus is the study of various CFTs with defects (dCFTs). 

The setting for these notes is $\mathcal{N}=4$ SYM with a codimension-one defect residing at the coordinate value $z=0$. 
The theory is the field theory dual of the probe D5-D3-brane system in $AdS_5\times S^5$, in which the probe-D5-brane has a three-dimensional intersection (the defect) with a stack of $N$ D3-branes. We will study the dual of the so called fuzzy-funnel solution\cite{ci:Karch:2000gx,ci:NAHM1980413,ci:Diaconescu:1996rk,ci:Constable:1999ac}, in which a background gauge field has $k$ units of flux through an $S^2$-part of the D5-brane geometry, meaning that $k$ D3-branes dissolve into the D5-brane. These parameters appear on the field theory side as the rank $N$ of the gauge group which is broken down to $N-k$ by the defect. 
\medskip

The dCFT action is built out of the regular $\mathcal{N}=4$ SYM field content plus additional fields constrained to the three dimensional defect. These additional fields interact both within themselves and with the bulk\footnote{meaning the region $z>0$} fields. However, only the six scalars from $\mathcal{N}=4$ SYM will play a role within these notes.

The defect breaks the 4D conformal symmetry down to those transformations that leave the boundary intact (\ie that map $z=0$ onto itself). Its  presence thus changes many of the general statements about CFTs, such as allowing for non-vanishing one-point functions and two-point functions between operators of different conformal dimensions. 
These new features were first studied in \cite{ci:Nagasaki:2012re,ci:Kristjansen:2012tn} and within the described setting, they have been the topic of a series of recent works.  Tree-level one-point functions in the \SU{2}- and \SU{3}-subsectors where considered in \cite{ci:deLeeuw:2015hxa,ci:BuhlMortensen:2015gfd,ci:deLeeuw:2016umh} while bulk propagators and loop corrections to the one-point functions where worked out in \cite{ci:BuhlMortensen:2016pxs,ci:BuhlMortensen:2016jqo,ci:BuhlMortensen:2017ind}. Two-point functions were very recently addressed in \cite{ci:deLeeuw:2017dkd} and earlier in \cite{ci:Liendo:2016ymz}.%
\footnote{Wilson loops in these settings with a defect have also attracted attention, see \eg \cite{ci:Nagasaki:2011ue,ci:Aguilera-Damia:2016bqv,ci:deLeeuw:2016vgp}.}

\medskip

The underlaying idea of all this business is to interpret single-trace operators as states in a spin-chain and employ the Bethe ansatz from within this context. The one-point functions were in this spirit found to be expressible in a compact determinant formula, making use of a special spin-chain state, called the Matrix Product State (\MPS), and Gaudin norm for Bethe states. The end result for the  tree-level one-point functions of operators
\begin{equation*}
  \op_L \sim \tr \Big(\overbrace{Z \dots Z X Z \dots Z X Z \dots}^{\text{$L$ complex scalars out of $M$ are $X$}} \Big)
\end{equation*}
in the \SU{2}-subsector was
\begin{align*}
  \expval{\op_L}_{\tree} = \frac{2^{L-1}}{z^L}
  C_2\left(\mathbf{u}\right) 
\sum_{j=\frac{1-k}{2}}^{\frac{k-1}{2}}  j^L \prod_{i=1}^{\frac{M}{2}} 
\frac{u_i^2\left(u_i^2 + \frac{k^2}{4}\right)}{\left[u_i^2+(j-\tfrac{1}{2})^2\right]
\left[u_i^2+(j+\tfrac{1}{2})^2\right]}
,
\end{align*}
under the condition that both the length $L$ and the number of excitations $M$ are even and that the set of $M$ Bethe rapidities has the special form $\mathbf{u}=\{u_1, -u_1, u_2, - u_2, \dots\}$. The parameter $k$ can be any positive integer and
\begin{equation*}
  C_2 \left(\mathbf{u}\right) =2\left[
 \left(\frac{2\pi ^2}{\lambda }\right)^L\frac{1}{L}
 \prod_{j}^{}\frac{u_j^2+\frac{1}{4}}{u_j^2}\,\,\frac{\det G^+}{\det G^-}\right]^{\frac{1}{2}}
 ,
\end{equation*} 
where $G^\pm$ are $\frac{M}{2}\times \frac{M}{2}$ matrices with matrix elements
\begin{equation*}
 G^\pm_{jk}=\left(\frac{L}{u_j^2+\frac{1}{4}}-\sum_{n}^{}K^+_{jn}\right)\delta _{jk}
 +K^\pm_{jk},
\end{equation*}
within which, in turn, 
\begin{equation*}
 K^\pm_{jk}=\frac{2}{1+\left(u_j-u_k\right)^2}\pm
 \frac{2}{1+\left(u_j+u_k\right)^2}\, .
\end{equation*}
The expression for $C_2$ was obtained from the spin-chain overlap
\begin{equation*}
  C_2 = \left( \frac{8\pi^2}{\lambda} \right)^{L/2} \! \frac{1}{\sqrt{L}} \,\frac{\braket{\MPS}{\Psi}}{\sqrt{ \braket{\Psi}{\Psi}}}
\end{equation*}
which is the form we will mostly refer to here. $\ket{\Psi}$ is the spin-chain Bethe state corresponding to the operator $\op_L$; the MPS will be defined below in equation \eqref{eq:defMPS}.

\subsection{\underline{The goal of the present notes}}
These notes consider the leading contribution, in the 't Hooft coupling $\lambda$, to the specific two-point function $\expval{\op_L \, \op_2}_{\onecontr}$, where 
\begin{itemize}[itemsep=.1ex,label=\raisebox{0.25ex}{\footnotesize$\bullet$}]
  \item both $\op_L$ and $\op_2$ are single-trace scalar operators of length $L$ and $2$, respectively, and 
  \item $\op_L$ is restricted to the \SU{2}-subsector while $\op_2$ can be built out of any pair of scalars. 
\end{itemize}
We do this by interpreting the contraction as a spin-chain operator $\QQ$ acting on the Bethe state corresponding to $\op_L$, whence re-expressing the two-point function in terms of the previously known one-point functions.

\section{The particular two-point functions}
We define the complex scalar fields as
\begin{align*}
  Z &= \phi_1 + \I \phi_4 \,,  	&	X &= \phi_2 + \I \phi_5 \,, 	&	W &= \phi_3 + \I \phi_6 \,,
  \\
  \Zb &= \phi_1 - \I \phi_4 \,,  	&	\Xb &= \phi_2 - \I \phi_5 \,, 	&	\Wb &= \phi_3 - \I \phi_6 \,,
\end{align*}
which in the dual fuzzy-funnel solution each has the non-zero classical expectation value
\begin{equation*}
  \phi_I^{\text{cl}} = \frac{1}{z} t_I \oplus \fatZero_{(N-k)}  , \quad I = 1, 2, 3; \qquad \phi_{\Jt}^{\text{cl}} = 0, \quad \Jt= 4, 5, 6,
\end{equation*}
where $\{t_1, t_2, t_3\}$ forms a $k\times k$ unitary representation of \SU{2} and the $\fatZero_{(N-k)}$ pads the rest of the matrix to the full dimensions $N\times N$.

For definiteness, we choose $Z\sim \ket{\ua}$ and $X\sim \ket{\da}$ as the \SU{2}-subsector.

We now set out to calculate 
\begin{align}
  \acontraction{%
    \expval{\op_L \op_{Y_1 Y_2 }}_{\text{1 contr.}} = \sum_{l = 1}^{L} \Psi^{i_1 \ldots i_L} \tr\big( X_{i_1}^{\cl} \cdots
  }{X}{%
    {}_{i_{l}}\cdots X_{i_L}^{\cl} \big) \tr\big( 
  }{Y}
  \expval{\op_L \op_{Y_1 Y_2} }_{\onecontr} &= \sum_{l = 1}^{L} \Psi^{i_1 \ldots i_L}
  \tr\big( X_{i_1}^{\cl} \cdots X_{i_{l}} \cdots X_{i_L}^{\cl} \big) \tr\big( Y_1 Y_2^{\cl} \big)
  \notag
  \\
  & \qquad + (Y_1 \leftrightarrow Y_2), \qquad i_{\ell} = \ua, \da
  \label{eq:corrFun}
\end{align}
where $X_{\ua}=Z$, $X_{\da} = X$, $Y_{1,2}$ can be any complex scalar and the coefficients $\Psi^{i_1 \dots i_L}$ of $\op_L$ are chosen such that they map to a Bethe state $\ket{\Psi}$ in the spin-chain picture. 

We will express it by help of the MPS, which is the following state in the spin-chain Hilbert space:
\begin{equation} \label{eq:defMPS}
  \bMPS = \tr \bigg[ \Big( \bra{\ua} t_1 + \bra{\da} t_2 \Big)^{\otimes L} \bigg]
  ,
\end{equation}
where the trace is over the resulting product of $t$'s.

\subsection{\underline{Scalar propagators}}
The defect mixes the scalar propagator in both color and flavor indices, explained in detail in \cite{ci:BuhlMortensen:2016jqo}. However, since the contracted fields are multiplied by classical fields from both sides we will only need the upper $(k\times k)$-block. The propagator diagonalization involves a decomposition of these components in terms of fuzzy spherical harmonics $\Y^m_{\ell}\,$:
\footnote{See appendices in \cite{ci:BuhlMortensen:2016jqo,ci:Kawamoto:2015qla}. We use the normalization of \cite{ci:BuhlMortensen:2016jqo}.}
\begin{equation*}
  [\phi]\indices{^{s_1}_{s_2}} = \sum_{\ell = 1}^{k-1} \sum_{m = -\ell}^{\ell} \phi_{\ell, m} [\Y^m_{\ell}]\indices{^{s_1}_{s_2}}, \qquad s_{1,2}=1, \dots, k
  .
\end{equation*}
Translating back to the $s$-indices, the relevant propagators for $I, J = 1, 2, 3$ read
\begin{align*}
  \expval{[\phi_I(x)]\indices{^{s_1}_{s_2}} [\phi_J(y)]\indices{^{r_1}_{r_2}}}
  =
  \delta_{I,J} \sum_{\ell, m} [\Y_{\ell}^m]\indices{^{s_1}_{s_2}} \, [(\Y_{\ell}^m)^\dagger]\indices{^{r_1}_{r_2}} \; K_{1}^{\ell}(x, y)
  \notag
  \\
  - \I \epsilon_{IJK}  \sum_{\ell, m, m'} [\Y_{\ell}^m]\indices{^{s_1}_{s_2}} \, [(\Y_{\ell}^{m'})^\dagger]\indices{^{r_1}_{r_2}} \; [t^{(2\ell+1)}_K ]_{\ell - m+1, \ell - m' +1} \; K_{2}^{\ell}(x, y)
\end{align*}
where $t^{(2\ell+1)}_K$ is in the $(2\ell+1)$-dimensional representation. The remaining scalars $\It, \Jt = 4, 5, 6$ have the diagonal propagator
\begin{equation*}
   \expval{[\phi_{\It}]\indices{^{s_1}_{s_2}} [\phi_{\Jt}]\indices{^{r_1}_{r_2}}}
   = \delta_{\It \Jt} \sum_{m=-\ell}^{\ell} [\Y_{\ell}^m]\indices{^{s_1}_{s_2}} \, [(\Y_{\ell}^m)^\dagger]\indices{^{r_1}_{r_2}} \; K^{\m^2=\ell(\ell+1)}(x, y)
   .
\end{equation*}
The spacetime dependent factors are
\begin{align*}
  K_{1}^{\ell}(x, y) & = \frac{\ell + 1}{2 \ell + 1} K^{\m^2=\ell(\ell-1)}(x, y) + \frac{\ell}{2\ell +1} K^{\m^2=(\ell + 1)(\ell+2)}(x, y)
  \,,
  \\
  K_{2}^{\ell}(x, y) &= \frac{1}{2\ell + 1} \left( K^{\m^2=\ell(\ell-1)}(x, y) - K^{\m^2=(\ell+1)(\ell+2)}(x, y) \right)
  .
\end{align*}
$K^{\m^2}$ is related to the scalar propagator in $AdS$ and reads
\begin{equation*}
  K^{\m^2}(x, y) = \frac{\gym^2}{2} (x_3 y_3)^{1/2} 
  \int\! \frac{\rd^3 \vec{k}}{(2\pi)^3} \; \e^{\I \vec{k}\cdot(\vec{x}-\vec{y})} \; I_\nu \big(|\vec{k}| x_3^<\big) \, K_\nu\big(|\vec{k}| x_3^> \big)
  ,
\end{equation*}
in which $I$ and $K$ are modified Bessel functions with $x_3^<$ ($x_3^>$) the smaller (larger) of $x_3$ and $y_3$, and lastly where $\nu = \sqrt{\m^2 + \tfrac{1}{4}}$. 

We will from now on suppress all spacetime dependence.

\section{The contraction as a spin-chain operator}
With the expressions of the propagators, we can now view the contraction in equation \eqref{eq:corrFun} as a $(k\times k)$-matrix
\begin{equation*}
  [\TT_{X_{i_l} Y_1 Y_2}]\indices{^{s_1}_{s_2}} = \expval{[X_{i_l}]\indices{^{s_1}_{s_2}} [Y_1]\indices{^{r_1}_{r_2}}} [\Ycl_2]\indices{^{r_2}_{r_1}}
\end{equation*}
replacing the field at site $l$ in the first trace while absorbing the second trace completely.

It turns out that this matrix always is proportional to either $t_1, t_2$ or $t_3$. To see this, first use that the fuzzy spherical harmonics are tensor operators, such that
\begin{equation*}
  \sum_m \Y^m_{\ell} [t^{(2\ell + 1)}_K]_{\ell - m +1, \ell - m' +1} = [t^{(k)}_{K}, \Y^{m'}_{\ell} ] = m' \Y^{m'}_{\ell}
  .
\end{equation*}
Then use the orthogonality of the fuzzy spherical harmonics%
\footnote{$\tr \Y^m_{\ell} (\Y^{m'}_{\ell'})^{\dagger} = \delta_{\ell \ell'} \delta_{m m'}\,.$}
 in the trace by decomposing the $t$ in $\Ycl_2$ as
\begin{align*}
  t_j &= d_j \, \big(\Y^{-1}_1 + (-1)^j \Y^1_1 \big),& \qquad &j=1 ,2,
  \\
  t_3 &= \sqrt{2} \, d_1 \Y^0_1 \,,		&
  &d_j = \I^{3+j} \frac{(-1)^{k+1}}{2} \sqrt{k(k^2-1)/6}
  .
\end{align*}
Together, these factors in $\TT$ then conspire to always give $t$'s for any considered scalar combination. What is left can thus be interpreted as a one-point function of a slightly modified $\op_L$. As such, we can write the two-point function \eqref{eq:corrFun} as an operator insertion 
\begin{equation*}
  \bMPS \QQ_{Y_1 Y_2} \ket{\Psi}
\end{equation*}
in the spin-chain picture, acting on the Bethe state corresponding to $\op_L$.

\subsection{\underline{The spin-chain operator} $\QQ_{Y_1 Y_2}$}
$\TT$'s dependence on the involved scalars can be compactly written when expanded in terms of the real scalars:
\begin{equation*}
  \TT_{IJK} = \delta^3_{IJ} K_1^{\ell=1} t_K + (\delta^3_{IK} t_J - \delta^3_{JK}t_I) K_2^{\ell=1} + \delta^6_{IJ} K^{\m^2=2} t_K
  ,
\end{equation*}
$I, J, K = 1,\dots, 6$ and where the $\delta^3$ $(\delta^6)$ is only non-zero for indices 1,2 and 3 (4, 5, and 6). Taking into account both the sums in the two-point function \eqref{eq:corrFun}, we can then write the contractions in the spin-chain picture as
\begin{equation*}
  \QQ_{Y_1 Y_2} \ket{\Psi} = \sum_{l=1}^{L} \eye \otimes \cdots \otimes \QQ^{(l)}_{Y_1 Y_2} \otimes \cdots \otimes \eye	\ket{\Psi}
  \,,
\end{equation*}
\ie a linear combination of the spin-chain operators $\{\eye^{\otimes L}, S^+, S^-, S^3\}$%
\footnote{This does not explicitly cover the case of $\TT \propto t_3$. However, that case eventually yields zero and will be addressed below.}
. 

The result arranges itself in the two cases $\Ycl_1 = \Ycl_2$ and $\Ycl_1 \neq \Ycl_2$, for which%
\footnote{We will denote both the dCFT operator and its spin-chain correspondent with subscripts ${}_=$ and ${}_{\neq}$ for these two cases.}
\begin{align*}
  \QQ^{(l)}_{=} &= 
  \begin{pmatrix}
    c^{\ua}	&	0	\\
    0	&	c^{\da}	\\
  \end{pmatrix}
  ,
  &
  \QQ^{(l)}_{\neq} &= 
  \begin{pmatrix}
    0	&	c^{+}	\\
    c^{-}	&	0	\\
  \end{pmatrix}
  ,
\end{align*}
and the various coefficients $c$ implicitly depend on $Y_1, Y_2$. They are listed in \ref{sec:coefList}.

\paragraph{\hspace{1ex}\textbullet{} Case $\Ycl_1 = \Ycl_2$}
The action of $\QQ_{=}$ is trivial on any Bethe state. Still denoting the total number of spin-down excitations as $M$, we immediately get
\begin{equation*}
  \QQ_{=} \ket{\Psi} = \big( c^{\ua} (L-M) + c^{\da} M \big) \ket{\Psi}
  .
\end{equation*}
Combining this with the one-point function formula implies
\begin{equation*}
  \expval{\op_L \op_{=}}_{\onecontr} = \big( c^{\ua} (L-M) + c^{\da} M \big) \expval{\op_L}_{\tree}
  .
\end{equation*}

As an example, the Konishi operator has the two-point function $2 K^{\m^2=6} L \expval{\op_L}_{\tree}$ with any \SU{2}-subsector operator.

\paragraph{\hspace{1ex}\textbullet{} Case $\Ycl_1 \neq \Ycl_2$}
In this case we have the spin-flipping operator
\begin{equation*}
  \QQ_{\neq} = c^+ S^+ + c^- S^-
  .
\end{equation*}
Its action simplifies significantly when acting on a Bethe state. First of all, Bethe states with non-zero momenta are highest weight states implying that $S^+ \ket{\Psi} = 0$. Secondly, we have that
\begin{equation*}
  S^- \ket{\Psi_M} = \lim_{p_{M+1} \to 0} \ket{\Psi_{M+1}}
  ,
\end{equation*}
meaning that acting on a Bethe state with the lowering operator creates a new Bethe state with one more excitation but with the corresponding momentum $p_{M+1}=0$. All other momenta are the same. These states are called (Bethe) descendants.

It was shown in \cite{ci:deLeeuw:2015hxa} that only states with $L$ and $M$ both even can have a non-zero overlap with the MPS. Furthermore, by studying the action of $Q_3$, the third conserved charge in the integrable hierarchy, it was proven that only unpaired%
\footnote{ ``unpaired'' refers to states which are invariant under parity transformation, implying momenta of the form $\{p_1, -p_1, \cdots \}$. 
}
states yield finite overlaps. This is true since $ Q_3 \kMPS = 0 $ and because $Q_3$ is non-zero on states that are not invariant under parity.

That $\QQ_{\neq}$ alters the number of excitations now makes it possible to have non-zero overlaps with states with odd $M$. However, since
\begin{equation*}
  [Q_3, S^-] = 0
\end{equation*}
the requirement of an unpaired state is still imposed. Hence, the only possible way for the overlap
\begin{equation*}
  \bMPS Q_{\neq} \ket{\Psi_{\Modd}}
\end{equation*}
to be non-vanishing is that that $\Modd$ is \emph{odd} and that the Bethe state is a descendant.

The general expression for such a state is
\begin{equation*}
  \ket{\Psi_{\Modd = M+n}} = (S^{-})^n \ket{\Psi_{M}}
  , \quad \text{$n$ odd.}
\end{equation*}
The two-point function \eqref{eq:corrFun} then follows from the commutation relation of the spin-operators, the action of $(S^-)^n$ on the MPS and the norm of the descendants\cite{ci:deLeeuw:2017dkd,ci:Escobedo:2010xs}:
\begin{align*}
  \bMPS (S^-)^n \ket{\Psi_{M}} &= \frac{n! (\tfrac{L}{2}-M)!}{\left( \frac{n}{2} \right)!  \left( \frac{L-2M-n}{2} \right)!} \braket{\MPS}{\Psi_{M}}
  \,,
  \\
  \braket{\Psi_{M+n}}{\Psi_{M+n}} &= \frac{n! (L-2M)!}{(L-2M-n)!} \braket{\Psi_{M}}{\Psi_M}
 .
\end{align*}
We find
\begin{multline*}
   \expval{\op_{L, M+n} \, \op_{\neq}}_{\onecontr} =
 \\
 \bigg(
  c^+  n \big( L-2M - n+1 \big) \, \mathcal{C}^+_{L, M, n} 
  + c^- \, \mathcal{C}^-_{L, M, n}
  \bigg)
  \expval{\op_{L, M}}_{\tree}
\end{multline*}
with 
\begin{equation}
  \mathcal{C}^{\pm}_{L, M, n} =
  \frac{ (n\mp 1)! \left(\frac{L}{2}-M \right)! }{\left( \frac{n\mp1}{2} \right)! \left( \frac{L-2M - n \pm 1}{2} \right)! } 
  \sqrt{\frac{(L-2M-n)!}{n! (L-2M)!}}
  .
  \label{eq:combFacs}
\end{equation}

\subsection{\underline{Remark on $\TT \propto t_3$}}
When \emph{one} of $Y_1$ or $Y_2$ is either $W$ or $\Wb$, $\TT$ is proportional to $t_3$ and the corresponding $\QQ^{(l)}_{t_3}$ is no longer a proper spin-chain operator. Insisting on a spin-chain interpretation would describe it as a flip of site $l+1$ followed by a removal of the site $l$, thus shrinking the length $L$ by one. $\QQ^{(l)}_{t_3}$ always appears preceded by a projection $\Pi_{\ua (\da)}$ on either spin-up or spin-down, depending on the $Y$ which does not involve $W (\Wb)$. It is straight-forward to show by explicit calculation that 
\begin{equation*}
  \bra{\MPS_{L-1}}\, \sum_{l=1}^L \eye \otimes \dots \otimes \QQ^{(l)}_{t_3} \Pi_{\ua(\da)}^{(l)} \otimes \dots \otimes \eye \ket{\updownarrow_L} = 0
\end{equation*}
for any basis vector $\ket{\updownarrow_L}$ of length $L$.

\section{Conclusion}
We have studied the $\mathcal{N}=4$ SYM theory with a defect, dual to the probe D5-D3-brane system. Within this theory, the two-point function  between a length $L$ operator $\op_L$ in the \SU{2}-subsector and any operator $\op_{Y_1 Y_2}$ of two scalars can, in the leading order, be written as a spin-chain operator insertion in the scalar product between a matrix product state $\bMPS$ and the Bethe state $\ket{\Psi}$ corresponding to the operator $\op_L$, 
\begin{equation*}
  \expval{\op_L \op_{Y_1 Y_2}}_{\onecontr} \propto \bMPS \QQ_{Y_1 Y_2} \ket{\Psi}
  .
\end{equation*}
The operation of $\QQ$ depends on the two fields $Y_1, Y_2$ but is simple for any choice of scalar fields:
\begin{itemize}[label=\raisebox{0.25ex}{\footnotesize$\bullet$}]
  \item For $Y^{\cl}_1 = Y^{\cl}_2$ we get
    \begin{equation*}
      \expval{\op_L \op_{Y_1 Y_2}} = \big(c^{\ua}L + c^{\da} (L-M) \big) \expval{\op_L}_{\tree}
    \end{equation*}
    where both $L$ and the number of excitations $M$ need to be \emph{even} and the Bethe state needs to be unpaired. 

  \item For $Y^{\cl}_1 \neq Y^{\cl}_2$, the two-point function is zero for any $\op_L$ mapping to a highest weight Bethe state. For operators $\op_{L, M+n}$ mapping to (Bethe) descendants, however, the two-point function is non-vanishing, under the condition that $n$ is odd and that the corresponding Bethe state descends from an unpaired state $\ket{\Psi_{L, M}}$. The result is
    \begin{multline*}
      \expval{\op_{L, M+n} \, \op_{Y_1 Y_2}}_{\onecontr} = 
      \\
     \bigg(
      c^+  n \big( L-2M - n+1 \big) \, \mathcal{C}^+_{L, M, n} 
      + c^- \, \mathcal{C}^-_{L, M, n}
      \bigg)
      \expval{\op_{L, M}}_{\tree}
      ,
    \end{multline*}
\end{itemize}
where the combinatorial factors $\mathcal{C}^{\pm}_{L, M, n}$ can be found in equation \eqref{eq:combFacs}.

The coefficients $c$ with various indices depend on $Y_1, Y_2$ and are all spacetime-dependent since they contain expressions of the propagator. See \ref{sec:coefList} below for the full list of coefficients.

These results hold for any $k$.

\section*{Acknowledgments}
We would like to thank  I. Buhl-Mortenssen, M. de Leeuw, F. Levkovich-Maslyuk, O. Ohlsson Sax, M. Wilhelm and K. Zarembo  for guidance and fruitful discussions.

This work was supported by the ERC advanced grant No 341222.

\appendix
\section{List of coefficients}\label{sec:coefList}
Here follows the list of coefficients for the considered two-point functions, written in the form $\QQ_{Y_1 Y_2}: \left( \begin{smallmatrix} c^{\ua} & c^+\\ c^{-} & c^{\da} \end{smallmatrix} \right)$.
  {\footnotesize
\begin{align*}
 \QQ_{Z Z} &:\quad \left(
\begin{array}{cc}
 \frac{2}{3} (2 K^{m^2=0}-3 K^{m^2=2}+K^{m^2=6}) & 0 \\
 0 & -\frac{2}{3} (K^{m^2=0}-K^{m^2=6}) \\
\end{array}
\right) \\[1ex]
 \QQ_{Z \Zb} &:\quad \left(
\begin{array}{cc}
 \frac{2}{3} (2 K^{m^2=0}+K^{m^2=6}) & 0 \\
 0 & -\frac{2}{3} (K^{m^2=0}-K^{m^2=6}) \\
\end{array}
\right) \\[1ex]
 \QQ_{\Zb \Zb} &:\quad \left(
\begin{array}{cc}
 \frac{2}{3} (2 K^{m^2=0}+3 K^{m^2=2}+K^{m^2=6}) & 0 \\
 0 & -\frac{2}{3} (K^{m^2=0}-K^{m^2=6}) \\
\end{array}
\right) \\[1ex]
 \QQ_{Z X} &:\quad \left(
\begin{array}{cc}
 0 & K^{m^2=0}-K^{m^2=2} \\
 K^{m^2=0}-K^{m^2=2} & 0 \\
\end{array}
\right) \\[1ex]
 \QQ_{Z \Xb} &:\quad \left(
\begin{array}{cc}
 0 & K^{m^2=0}+K^{m^2=2} \\
 K^{m^2=0}-K^{m^2=2} & 0 \\
\end{array}
\right) \\[1ex]
 \QQ_{\Zb X} &:\quad \left(
\begin{array}{cc}
 0 & K^{m^2=0}-K^{m^2=2} \\
 K^{m^2=0}+K^{m^2=2} & 0 \\
\end{array}
\right) \\[1ex]
 \QQ_{\Zb \Xb} &:\quad \left(
\begin{array}{cc}
 0 & K^{m^2=0}+K^{m^2=2} \\
 K^{m^2=0}+K^{m^2=2} & 0 \\
\end{array}
\right) \\[1ex]
 \QQ_{X X} &:\quad \left(
\begin{array}{cc}
 -\frac{2}{3} (K^{m^2=0}-K^{m^2=6}) & 0 \\
 0 & \frac{2}{3} (2 K^{m^2=0}-3 K^{m^2=2}+K^{m^2=6}) \\
\end{array}
\right) \\[1ex]
 \QQ_{X \Xb} &:\quad \left(
\begin{array}{cc}
 -\frac{2}{3} (K^{m^2=0}-K^{m^2=6}) & 0 \\
 0 & \frac{2}{3} (2 K^{m^2=0}+K^{m^2=6}) \\
\end{array}
\right) \\[1ex]
 \QQ_{\Xb \Xb} &:\quad \left(
\begin{array}{cc}
 -\frac{2}{3} (K^{m^2=0}-K^{m^2=6}) & 0 \\
 0 & \frac{2}{3} (2 K^{m^2=0}+3 K^{m^2=2}+K^{m^2=6}) \\
\end{array}
\right) \\[1ex]
\QQ_{W W} &= \QQ_{W \Wb} = \QQ_{\Wb \Wb} : \left(
\begin{array}{cc}
 -\frac{2}{3} (K^{m^2=0}-K^{m^2=6}) & 0 \\
 0 & -\frac{2}{3} (K^{m^2=0}-K^{m^2=6}) \\
\end{array}
\right) \\[1ex]
\end{align*}
}


\makebox[0pt]{\raisebox{-4ex}[0pt][0pt]{\color{white}$\varheartsuit \, 1+1=3 \,\varheartsuit$}}

\end{document}